\documentclass[a4paper,final]{appolb}
 \usepackage[]{graphicx} 
\usepackage{showkeys}
\begin{document}
%
\pagestyle{plain}
\title{
Highly anisotropic hydrodynamics -- discussion of the model assumptions and forms of the initial conditions \footnote{Supported in part by the Polish Ministry of Science and Higher Education, grants  \newline N N202 288638 and N  N202 263438.}}
\author{Rados{\l}aw Ryblewski $^{1}$ and  Wojciech Florkowski $^{2,1}$  
\address{
$^1$ H. Niewodnicza\'nski Institute of Nuclear Physics, Polish Academy of Sciences, PL-31342 Krak\'ow, Poland \\
$^2$ Institute of Physics, Jan Kochanowski University, PL-25406~Kielce,~Poland
}}

\date{November 29, 2010}
\maketitle
\begin{abstract}
The model assumptions of the recently formulated framework of highly-anisotropic and strongly-dissipative hydrodynamics (ADHYDRO) are analyzed. In particular, we study dependence of numerical results on different forms of the entropy source and compare our approach with other frameworks describing locally anisotropic fluids. We also discuss the effects of different forms of the initial conditions on the process of isotropization.
\end{abstract}
%
\section{Introduction}
\label{sect:introduction}

Recently, we have introduced the framework of highly-anisotropic and strongly-dissipative hydrodynamics \cite{Florkowski:2010cf,Ryblewski:2010bs} (ADHYDRO). This framework is suitable for modeling of early stages of relativistic heavy-ion collisions, in the situations where very large pressure anisotropies are expected \cite{Kovner:1995ja,Bjoraker:2000cf,El:2007vg}. In Ref. \cite{Florkowski:2010cf} our approach was used to describe boost-invariant, one-dimensional  systems, whereas in Ref. \cite{Ryblewski:2010bs} we applied ADHYDRO in non-boost-invariant situations. 

ADHYDRO is a simple extension of the perfect-fluid hydrodynamics. For locally isotropic fluids it is reduced to perfect-fluid description, and for small pressure anisotropies and purely longitudinal expansion  it agrees with the 2nd order Israel-Stewart viscous hydrodynamics (we note that longitudinal expansion dominates at the early stages of heavy-ion collisions). For highly-anisotropic systems, ADHYDRO may be treated as an effective model which describes strongly non-equilibrium dynamics. 

The construction of ADHYDRO was motivated by the idea of developing an approach that is suitable for uniform  description of several stages of the evolution of matter --- starting from the very early stages where the momentum anisotropies are extremely high, through the transient viscous evolution, and ending with the perfect-fluid stage. Since most of the realistic hydrodynamic calculations indicate that the evolution of the system is very well described by the perfect-fluid hydrodynamics (or, equivalently, by viscous hydrodynamics with a very small ratio of the shear viscosity to entropy) with the starting proper time below \mbox{$\tau_{\rm hyd} =$ 1 fm} ~\footnote{We use the natural units where $\hbar=c=1$.}   \cite{Huovinen:2001cy,Teaney:2001av,Kolb:2002ve,Broniowski:2008vp,Pratt:2008qv,Bozek:2009ty}, we expect that the ADHYDRO description is reduced or becomes very close to perfect-fluid hydrodynamics at the proper time $\tau \approx \tau_{\rm hyd}$. On the other hand, at the very early stages of the collisions, $\tau < \tau_{\rm hyd}$, neither perfect-fluid hydrodynamics nor viscous hydrodynamics are formally applicable. In particular, viscous corrections become compatible with the leading terms, which may lead to negative pressure. Thus, the very early dynamics requires the use of the kinetic theory (which has its own well known limitations) or modeling. ADHYDRO is a proposal for such modeling. 

A work along similar lines have been recently done by Martinez and Strickland \cite{Martinez:2010sc,Martinez:2010sd}, who derive analogous equations from the Boltzmann equation with the collision term treated in the relaxation time approximation. Instead of the usual expansion around the isotropic one-particle distribution, they consider an expansion around an anisotropic distribution. In this paper we compare some aspects of our approach with that presented in \cite{Martinez:2010sc,Martinez:2010sd}.

The main model assumptions of ADHYDRO are the form of the energy-momentum tensor that allows for anisotropic pressure and the form of the entropy source. For small anisotropies, the form of the entropy source  is restricted by the consistency with the 2nd order viscous hydrodynamics. On the other hand, for large anisotropies, various forms of the entropy source are conceivable. Of course it would be very useful to obtain some hints about the form of the entropy source from the underlying microscopic dynamics. Such connections are, however, difficult to find since the initial dynamics of the system produced in heavy-ion collisions is very much complicated. Therefore, in this paper we simply use different forms of the entropy source and study their effects on the evolution of the system. 

Besides the analysis of the entropy source term, we discuss the microscopic interpretation of our framework and analyze the effects of quantum statistics. We find that the inclusion of the Bose-Einstein statistics leads to minor quantitative modifications of the approach introduced in  \cite{Florkowski:2010cf}. Moreover, in this paper we present new results obtained for the initial pressure anisotropy that depends on rapidity. We demonstrate that such initial conditions do not affect the isotropization process. 

\section{Basic equations}
\label{sect:basic}

The ADHYDRO model \cite{Florkowski:2010cf}  is based on the following form of the energy-momentum tensor  \cite{Florkowski:2008ag,Florkowski:2009sw}
\begin{eqnarray}
T^{\mu \nu} &=& \left( \varepsilon  + P_\perp\right) U^{\mu}U^{\nu} 
- P_\perp \, g^{\mu\nu} - (P_\perp - P_\parallel) V^{\mu}V^{\nu}. 
\label{Tmunudec}
\end{eqnarray}
In Eq. (\ref{Tmunudec}) $\varepsilon$, $P_\perp$, and $P_\parallel$ are the energy density, transverse pressure, and longitudinal pressure, respectively. For \mbox{$P_\perp=P_\parallel=P$} we recover the form valid for the perfect fluid. The four-vector $U^\mu$ describes the hydrodynamic flow
\begin{equation}
U^\mu = \gamma (1, v_x, v_y, v_z), \quad \gamma = (1-v^2)^{-1/2},
\label{Umu}
\end{equation}
while $V^\mu$ defines the longitudinal direction (corresponding to the collision axis)
\begin{equation}
V^\mu = \gamma_z (v_z, 0, 0, 1), \quad \gamma_z = (1-v_z^2)^{-1/2}.
\label{Vmu}
\end{equation}
The four-vectors $U^\mu$ and $V^\mu$ satisfy the following normalization conditions:
\begin{eqnarray}
U^2 = 1, \quad V^2 = -1, \quad U \cdot V = 0.
\label{UVnorm}
\end{eqnarray}
In the local-rest-frame (LRF) of the fluid element the four-vectors $U^\mu$ and $V^\mu$ have simple forms,
\begin{eqnarray}
 U^\mu = (1,0,0,0), \quad V^\mu = (0,0,0,1), 
\end{eqnarray}
and the energy-momentum tensor has the following diagonal structure~\footnote{The Lorentz transformation leading to LRF is defined explicitly in Sect.~\ref{sect:app1}. Since this transformation should not change the longitudinal direction, it must be defined in a special way.}
\begin{equation}
T^{\mu \nu} =  \left(
\begin{array}{cccc}
\varepsilon & 0 & 0 & 0 \\
0 & P_\perp & 0 & 0 \\
0 & 0 & P_\perp & 0 \\
0 & 0 & 0 & P_\parallel
\end{array} \right).
\label{Tmunuarray}
\end{equation}

In addition to the energy-momentum tensor (\ref{Tmunudec}) we introduce the entropy flux
\begin{eqnarray}
\sigma^{\mu} &=& \sigma U^{\mu},
\label{smudec}
\end{eqnarray}
where $\sigma$ is the entropy density. In our approach $\varepsilon$ and  $\sigma$ are functions of the two pressures, $P_\perp$ and $P_\parallel$. In particular, for massless partons we have
\begin{equation}
\varepsilon = 2 P_\perp + P_\parallel.
\label{eos}
\end{equation}
The space-time dynamics of the system is described by the following equations
\begin{eqnarray}
\partial_\mu T^{\mu \nu} &=& 0, \label{enmomcon} \\
\partial_\mu \sigma^{\mu} &=& \Sigma. \label{engrow}
\end{eqnarray}
Equation  (\ref{enmomcon}) expresses the energy-momentum conservation, while Eq. (\ref{engrow}) describes the entropy production. The form of the entropy source $\Sigma$ determines the dynamics of the anisotropic system~\footnote{The case $\Sigma=0$ was analyzed in Refs. \cite{Florkowski:2008ag,Florkowski:2009sw}.}. This issue will be discussed in more detail below. 

If the functions $\sigma(P_\perp,P_\parallel)$ and $\Sigma(P_\perp,P_\parallel)$ are specified, Eqs. (\ref{enmomcon}) and (\ref{engrow}) form a closed system of 5 equations for 5 unknown functions: three components of the fluid velocity ${\bf v}$, the transverse pressure $P_\perp$, and the longitudinal pressure $P_\parallel$. The projections of Eq. (\ref{enmomcon}) on $U_\nu$ and $V_\nu$ yield \cite{Florkowski:2010cf}
\begin{eqnarray}
U^\mu \partial_\mu \varepsilon &=& - \left( \varepsilon+P_\perp \right) \partial_\mu U^\mu 
+ \left( P_\perp-P_\parallel \right) U_\nu  V^\mu \partial_\mu V^\nu, \label{enmomconU} \\
V^\mu \partial_\mu P_\parallel &=& - \left( P_\parallel-P_\perp \right) \partial_\mu V^\mu 
+ \left( \varepsilon+ P_\perp \right) V_\nu  U^\mu \partial_\mu U^\nu. \label{enmomconV}
\end{eqnarray}
For one-dimensional motion along the beam axis it is sufficient to consider Eqs. (\ref{engrow}), (\ref{enmomconU}), and (\ref{enmomconV}). Moreover, if the motion is boost-invariant, Eq. (\ref{enmomconV}) is satisfied automatically and we are left with Eqs. (\ref{engrow})  and  (\ref{enmomconU}) only.

\section{Microscopic interpretation}
\label{sect:aniso}

The anisotropic structure of the energy-momentum tensor appears naturally if partons are described by the phase space distribution function 
\begin{equation}
f = f\left( \frac{p_\perp}{\lambda_\perp},\frac{|p_\parallel|}{\lambda_\parallel}\right),
\label{Fxp1}
\end{equation}
where $\lambda_\perp$ and $\lambda_\parallel$ are two different space-time dependent scales that may be interpreted as transverse and longitudinal temperature, respectively. The form (\ref{Fxp1}) is valid in the local rest frame of the fluid element. The manifestly covariant form of $f$ is
\begin{equation}
f  = f\left( \frac{\sqrt{(p \cdot U)^2 - (p \cdot V)^2 }}{\lambda _\perp }, 
\frac{|p \cdot V|}{\lambda _\parallel  }\right).
\label{Fxp2}
\end{equation}
We note that as a special case of (\ref{Fxp2}) we obtain the formalism of transverse hydrodynamics \cite{Bialas:2007gn,Ryblewski:2008fx}, see Sect. \ref{sect:app2} for more details.

If the distribution function is known, one can calculate easily the particle density
\begin{eqnarray}
n &=& \int \frac{d^3p}{(2\pi)^3} \, 
f\left( \frac{p_\perp}{\lambda _\perp}, \frac{| p_\parallel |}{\lambda _\parallel }\right).
\label{rho1} 
\end{eqnarray}
Similarly, we calculate the energy density
\begin{eqnarray}
\varepsilon &=&   \int \frac{d^3p}{ (2\pi)^3}  \, E_p \, 
f\left( \frac{p_\perp}{\lambda _\perp}, \frac{| p_\parallel |}{\lambda _\parallel }\right),  
\label{epsilon1} 
\end{eqnarray}
where $E_p = \sqrt{p_\perp^2+p_\parallel^2}$\,, and the transverse and longitudinal pressures 
\begin{eqnarray}
P_\perp &=&   \int \frac{d^3p}{ (2\pi)^3}  \, \frac{p_\perp^2}{2 E_p} \, f\left( \frac{p_\perp}{\lambda _\perp}, \frac{| p_\parallel |}{\lambda _\parallel }\right),  \label{PT1} 
\end{eqnarray}
\begin{eqnarray}
P_\parallel &=&   \int \frac{d^3p}{ (2\pi)^3}  \, \frac{p_\parallel^2}{E_p} \, f\left( \frac{p_\perp}{\lambda _\perp}, \frac{| p_\parallel |}{\lambda _\parallel }\right).  \label{PL1} 
\end{eqnarray}

Equations (\ref{rho1})--(\ref{PL1}) indicate that we may use $\lambda_\perp$ and $\lambda_\parallel$ alternatively with $P_\perp$ and $P_\parallel$.  In Ref. \cite{Florkowski:2009sw} we have shown, however, that the two most convenient thermodynamics-like parameters are the particle density $n$ and the anisotropy parameter $x$ defined by the equation~\footnote{Since the entropy density is proportional to the particle number density, an alternative choice is to use the entropy density $\sigma$ and $x$, see Eqs. (\ref{vareCL})--(\ref{PLCL}).}
\begin{equation}
x = \left( \frac{\lambda_\perp}{\lambda_\parallel} \right)^2.
\end{equation}
With the help of those two variables we may express the energy density  and pressures in a compact form \cite{Florkowski:2008ag,Florkowski:2009sw}, namely 
\begin{eqnarray}
\varepsilon (x,n) &=& \left( \frac{n}{g} \right)^{4/3} R(x), 
\label{vareR} \\ 
P_{\perp} (x,n)   &=& \left( \frac{n}{g} \right)^{4/3} \left[ \frac{R(x)}{3} + x R'(x) \right], 
\label{PTR}\\
P_{\parallel}(x,n)&=& \left( \frac{n}{g} \right)^{4/3} \left[ \frac{R(x)}{3} - 2 x R'(x) \right], 
\label{PLR}
\end{eqnarray}
where the function $R(x)$ is defined by the integral
\begin{equation}
R(x) = x^{-1/3} \int \frac{d\xi_\perp\,\xi_\perp\,d\xi_\parallel}{2\pi^2}
\sqrt{\xi_\parallel^2 + x \xi_\perp^2} f(\xi_\perp,\xi_\parallel),
\label{Rofiks}
\end{equation}
and $g$ is a constant defined by the expression
\begin{equation}
g =  \int \frac{d\xi_\perp\,\xi_\perp\,d\xi_\parallel}{2\pi^2}
 f(\xi_\perp,\xi_\parallel).
\label{gconst}
\end{equation}
Here $\xi_\perp=p_\perp/\lambda_\perp$ and $\xi_\parallel=p_\parallel/\lambda_\parallel$ are two dimensionless parameters. The range of the integration over $\xi_\perp$ and $\xi_\parallel$ is always between 0 and infinity. The function $R^\prime(x)$ in Eqs. (\ref{PTR}) and (\ref{PLR}) is the derivative of $R(x)$ with respect to $x$. It is expected that $R'(x)$ vanishes at $x=1$, in this case the value $x=1$ corresponds to equilibrium where $P_\perp=P_\parallel$.

\section{Entropy source}
\label{sect:Sigma}

To solve Eqs. (\ref{enmomcon}) and  (\ref{engrow}) we have to assume a certain form of the entropy source, $\Sigma = \Sigma (\sigma, x)$. In Ref. \cite{Florkowski:2010cf} we proposed the following ansatz
\begin{equation}
\Sigma = \Sigma_0(\sigma,x) = \frac{(\lambda_{\perp}-\lambda_{\parallel})^{2}}{\lambda_{\perp} \lambda_{\parallel}}\frac{\sigma}{\tau_{\rm eq}} = \frac{(1-\sqrt{x})^{2}}{\sqrt{x}}\frac{\sigma}{\tau_{\rm eq}},
\label{en1}
\end{equation}
where $\tau_{\rm eq}$ is a time-scale parameter. The expression on the right-hand-side of Eq. (\ref{en1}) has several appealing features: i) it is positive, as expected on the grounds of the second law of thermodynamics, ii) it has a correct dimension, iii) it vanishes in equilibrium, where $x=1$, iv) it is symmetric with respect to the interchange of $\lambda_\perp$ with $\lambda_\parallel$ (consequently $\Sigma$ does not change if $x \to 1/x$), finally v) for small deviations from equilibrium, where $|x-1| \ll 1$, we find
\begin{equation}
\Sigma_0 (x) \approx \frac{(x-1)^{2}}{4 \tau_{\rm eq}} \sigma.
\label{en1exp}
\end{equation}
The quadratic dependence displayed in (\ref{en1exp}) is characteristic for the 2nd order viscous hydrodynamics~\footnote{For the 2nd order viscous hydrodynamics, the entropy production is proportional to the viscous stress squared, $\Sigma \propto \Pi^2$ \cite{Muronga:2003ta}. On the other hand, for small $x-1$ one can find that $\Pi$ is proportional to $x-1$ \cite{Martinez:2010sc}. See also Ref. \cite{Bozek:2007di}}. Below we consider also the entropy source of the form
\begin{equation}
\Sigma = {\tilde \Sigma} (\sigma,x) = \frac{(P_{\perp}-P_{\parallel})^{2}}{P_{\perp} P_{\parallel}}\frac{\sigma}{{\tilde \tau}_{\rm eq}},
\label{en2}
\end{equation}
where ${\tilde \tau}_{\rm eq}$ is another time-scale parameter that may be different from ${\tau}_{\rm eq}$. 

In addition to expressions (\ref{en1}) and (\ref{en2}) we also identify the entropy source in the model that has been recently proposed by Martinez and Strickland \cite{Martinez:2010sc}. This issue is discussed in Sect. \ref{subsect:MSmodel}.

\section{Standard formulation and corrections for quantum statistics}
\label{subsect:stand}

\subsection{Anisotropic Boltzmann distribution}

\begin{figure}[t]
\begin{center}
\includegraphics[angle=0,width=0.55\textwidth]{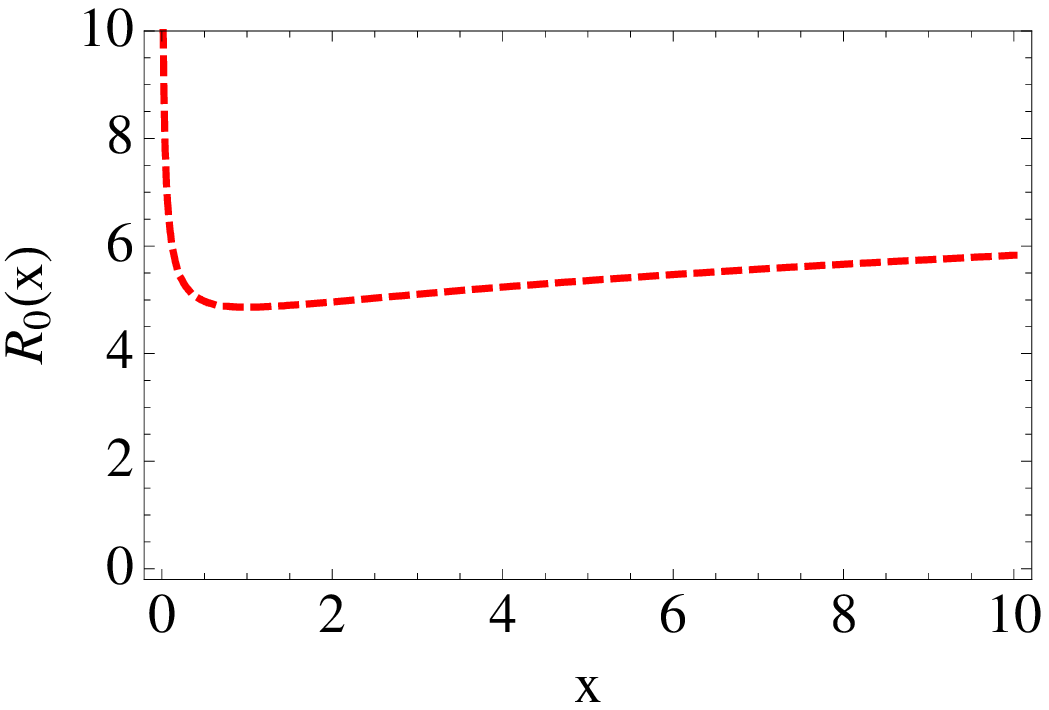} \\
\includegraphics[angle=0,width=0.55\textwidth]{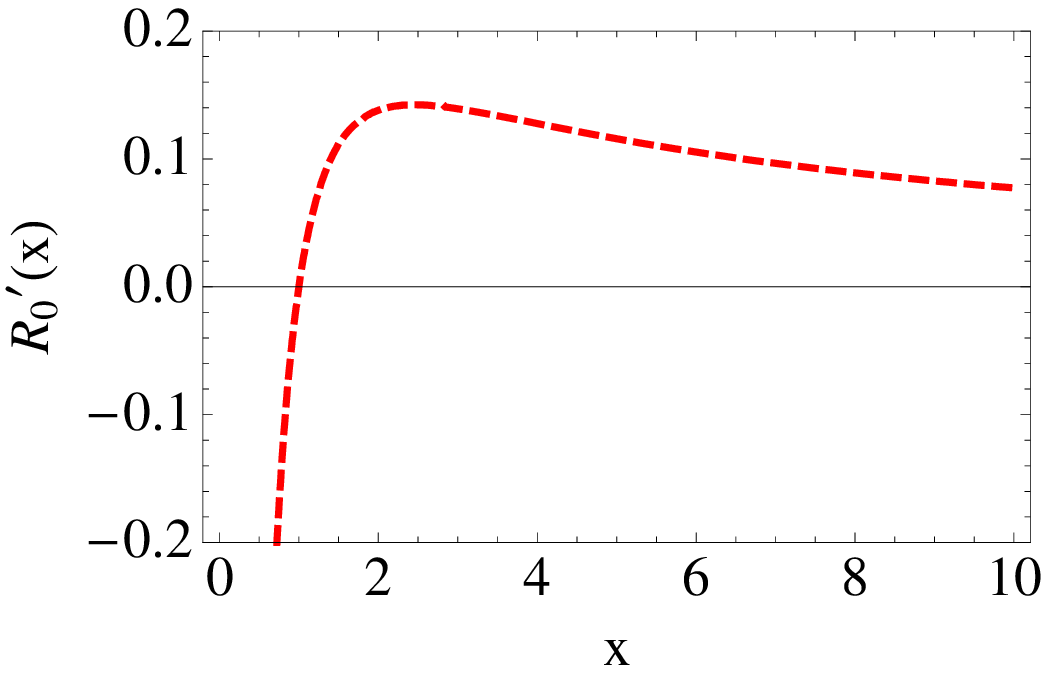} 
\end{center}
\caption{Function $R_0(x)$ defined by Eq. (\ref{ourR}) (upper part) and its derivative $R_0'(x)$ (lower part).}
\label{fig:RandRp}
\end{figure}

Before we switch to the discussion of the effects connected with the use of different forms of the entropy source, it is interesting to discuss different forms of the distribution function which motivates the structure of the energy-momentum tensor used in our approach. Our previous calculations have been based on the following exponential form of the distribution function 
\begin{equation}
f = g_0 \exp \left( -\sqrt{p_\perp ^2/\lambda_\perp^2 + p_\parallel^2/\lambda_\parallel^2 } \,  \right).
\label{fstand}
\end{equation}
Here $g_0$ is the degeneracy factor connected with internal quantum numbers such as spin, flavor,  or color. Since we assume that the system formed at the early stages of collisions consists mainly of gluons, we take $g_0 = 16$. Consistently with (\ref{fstand}), the entropy density $\sigma$ has been obtained from the Boltzmann non-equilibrium definition
\begin{eqnarray}
\sigma = \frac{g_0 \lambda_\perp^2 \,\lambda_\parallel}{2\pi^2} \int
d\xi_\perp\,\xi_\perp\,  d\xi_\parallel \, \frac{f\left(\xi_\perp,\xi_\parallel \right)}{g_0} \left[1 - \ln \frac{f\left(\xi_\perp,\xi_\parallel \right)}{g_0} \right].
\label{sigma}
\end{eqnarray}
Comparing (\ref{sigma}) with the particle density obtained from (\ref{rho1}) one gets
\begin{equation}
\sigma = 4 \,n.
\end{equation}
In this case the parameter $g$ obtained from Eq. (\ref{gconst}) equals $g_0/\pi^2$ and Eqs.~(\ref{vareR})--(\ref{PLR}) may be rewritten in the equivalent form as 
\begin{eqnarray}
\varepsilon (x,\sigma) &=& \left( \frac{\pi^2 \sigma}{4 g_0} \right)^{4/3} R_0(x), 
\label{vareCL} \\ 
P_{\perp} (x,\sigma)   &=& \left( \frac{\pi^2 \sigma}{4 g_0} \right)^{4/3} \left[ \frac{R_0(x)}{3} + x R_0'(x) \right], 
\label{PTCL}\\
P_{\parallel}(x,\sigma)&=& \left( \frac{\pi^2 \sigma}{4 g_0} \right)^{4/3} \left[ \frac{R_0(x)}{3} - 2 x R_0'(x) \right], 
\label{PLCL}
\end{eqnarray}
where $R_0(x)$ is the function obtained from Eq. (\ref{Rofiks}) with the distribution (\ref{fstand}), namely
\begin{equation}
R_0(x) = \frac{3\, g_0\, x^{-\frac{1}{3}}}{2 \pi^2} \left[ 1 + \frac{x \arctan\sqrt{x-1}}{\sqrt{x-1}}\right].
\label{ourR}
\end{equation}
The function $R_0(x)$ and its derivative are shown in Fig. \ref{fig:RandRp}, we note that $R_0(1) = 3 g_0/\pi^2$ and, as expected, $R_0^\prime(1)=0$.

\subsection{Anisotropic Bose-Einstein distribution}

The function (\ref{fstand}) may be considered as a Boltzmann equilibrium distribution that has been stretched (or squeezed) in the longitudinal direction. It is interesting to consider, in the similar way, the stretched (or squeezed) Bose-Einstein distribution, 
\begin{equation}
f _{BE} = \frac{g_0} {\exp \left( \sqrt{p_\perp ^2/\lambda_\perp  ^2 + 
p_\parallel^2/\lambda_\parallel^2}  \,  \right) - 1}.
\label{fBE}
\end{equation}
The corresponding definition of the entropy density for bosons is \cite{LandauSPh}
\begin{eqnarray}
\sigma = - \frac{g_0 \lambda_\perp^2 \,\lambda_\parallel}{2\pi^2} \int
d\xi_\perp\,\xi_\perp\,  d\xi_\parallel \, \left[
\frac{f_{BE}}{g_0} \ln \frac{f_{BE}}{g_0} -
\left(1+\frac{f_{BE}}{g_0}\right)  \ln \left(1+\frac{f_{BE}}{g_0}\right) \right]. \nonumber \\
\label{sigmaBE}
\end{eqnarray}
In this case the connection between the entropy density and the particle density is the same as in the isotropic gas of massless bosons, 
\begin{equation}
\sigma = \frac{2 \pi^4}{45 \zeta(3)} \, n.
\end{equation}
Substituting Eq. (\ref{fBE}) into Eqs. (\ref{Rofiks}) and (\ref{gconst}) we find, see Sect. \ref{sect:app3},
\begin{equation}
g_{BE} = \zeta(3) \frac{g_0}{\pi^2},
\end{equation}
\begin{equation}
R_{BE}(x) = \zeta(4) \, R_0(x) = \frac{\pi^4}{90} R_0(x).
\end{equation}
Thus, for quantum statistics we obtain
\begin{eqnarray}
\varepsilon (x,\sigma) &=& \left( \frac{45\, \zeta^{3/4}(4)\, \sigma}{2 \pi^2 g_0} \right)^{4/3}  R_0(x), 
\label{vareBE} \\ 
P_{\perp} (x,\sigma)   &=& \left( \frac{45\, \zeta^{3/4}(4)\, \sigma}{2 \pi^2 g_0} \right)^{4/3} 
\left[ \frac{R_0(x)}{3} + x R_0'(x) \right], 
\label{PTBE}\\
P_{\parallel}(x,\sigma)&=& \left( \frac{45\, \zeta^{3/4}(4)\, \sigma}{2 \pi^2 g_0} \right)^{4/3}  
\left[ \frac{R_0(x)}{3} - 2 x R_0'(x) \right].
\label{PLBE}
\end{eqnarray}
We note that the numerical factor $45 \, \zeta^{3/4}(4)/(2 \pi^2) \approx 2.42$ appearing in Eqs. (\ref{vareBE})--(\ref{PLBE}) is very close to the factor $\pi^2/4 \approx 2.47$ appearing in Eqs. (\ref{vareCL})--(\ref{PLCL}), hence, the equations defining the energy density and pressures in terms of the entropy density and $x$ for the classical and quantum systems are practically the same. 

\section{One-dimensional boost-invariant motion}
\label{sect:1dim}

In the case of the longitudinal and boost-invariant evolution, the four-vectors $U^\mu$ and $V^\mu$ may be defined by the expressions
\begin{eqnarray}
U^{\mu} = (\cosh \eta , 0 , 0 ,\sinh \eta), \label{fig:Umu} \\
V^{\mu} = (\sinh \eta , 0 , 0 ,\cosh \eta),\label{fig:Vmu}
\end{eqnarray}
where $\eta$ is the space-time rapidity,
\begin{eqnarray}
\eta = \frac{1}{2} \ln \frac{t+z}{t-z}. \label{eta} 
\end{eqnarray}
Substituting Eqs. (\ref{fig:Umu}) and (\ref{fig:Vmu}) in Eqs. (\ref{enmomcon}) and (\ref{engrow}) we obtain
\begin{eqnarray}
&& \frac{d \varepsilon}{d \tau} = -\frac{\varepsilon + P_\parallel }{\tau},
\label{eq1binv} \\
&& \frac{d \sigma}{d \tau} + \frac{\sigma}{\tau} = \Sigma,
\label{eq3binv}
\end{eqnarray}
where $\tau$ is the proper time
\begin{equation}
\tau = \sqrt{t^2 - z^2}.
\label{tau}
\end{equation} 

\subsection{Standard formulation}
\label{subsect:standard}

Substituting Eqs. (\ref{vareCL}) and (\ref{PLCL}) in  (\ref{eq1binv}), and using Eq. (\ref{en1}) for the entropy source, we obtain one ordinary differential equation for $x$,
\begin{equation}
\frac{dx}{d\tau} = \frac{2x}{\tau} -\frac{4 H_0(x)}{3 \tau_{\rm eq}}.
\label{xoftau0}
\end{equation}
We have introduced here the function $H_0(x)$ defined by the expression
\begin{eqnarray}
H_0(x)= \frac{R_0(x)}{R_0'(x)} \frac{\Sigma}{\sigma} \tau_{\rm eq} =
\frac{R_0(x)}{R_0'(x)} \frac{(1-\sqrt{x})^{2}}{\sqrt{x}} 
\label{H0}.
\end{eqnarray}
For small deviations from equilibrium we may use the expansion
\begin{eqnarray}
H_0(x)  \approx \frac{45 }{16}(x-1)+\frac{195}{112} (x-1)^2 + \cdots . 
\label{H0exp}
\end{eqnarray}
Equation (\ref{xoftau0}) was analyzed in detail in Ref. \cite{Florkowski:2010cf}, where we showed that $x \to 1$ for $\tau \gg \tau_{\rm eq}$. 

\subsection{Modified entropy source}
\label{subsect:modified}

Let us now turn to the discussion of the entropy source defined by the difference of pressures. If we use Eq. (\ref{en2}) instead of Eq. (\ref{en1}), we obtain

\begin{equation}
\frac{dx}{d\tau} = \frac{2x}{\tau} -\frac{4 {\tilde H}(x)}{3 {\tilde \tau}_{\rm eq}},
\label{xoftauT}
\end{equation}
where
\begin{eqnarray}
{\tilde H}(x)= 
\frac{9 x^2 R_0(x) R_0^\prime (x) }{\left( R_0(x)/3+x R_0^\prime(x) \right)\left(R_0(x)/3-2x R_0^\prime(x)\right)} 
\label{HT}.
\end{eqnarray}
Similarly to (\ref{H0exp}) we may expand (\ref{HT}) around $x=1$,
\begin{eqnarray}
\tilde{H}(x)  \approx \frac{36  }{5}(x-1)+\frac{816 }{175}(x-1) ^2+ \cdots .
\label{HTexp}
\end{eqnarray}
The requirement that the two definitions of the entropy source lead to the same behavior in the region close to equilibrium leads to the condition
\begin{equation}
{\tilde \tau}_{\rm eq} = \frac{64}{25} \,{\tau}_{\rm eq}.
\label{twotimes}
\end{equation}

\begin{figure}[t]
\begin{center}
\includegraphics[angle=0,width=0.65\textwidth]{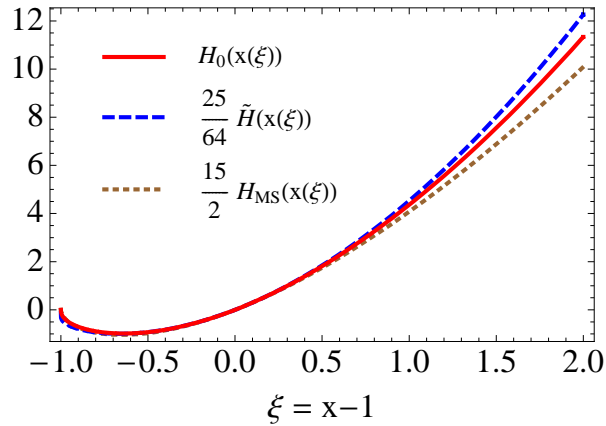}
\end{center}
\caption{Function $H_0$ defined by Eq. (\ref{H0}) (solid line) compared to the rescaled functions ${\tilde H}$ and  $H_{{\rm MS}}$ defined by Eqs. (\ref{HT}) and (\ref{Hms}) (dashed and dotted lines, respectively). The arguments of the functions have been shifted to $\xi=x-1$.}
\label{fig:H}
\end{figure}

\begin{figure}[t]
\begin{center}
\includegraphics[angle=0,width=0.65\textwidth]{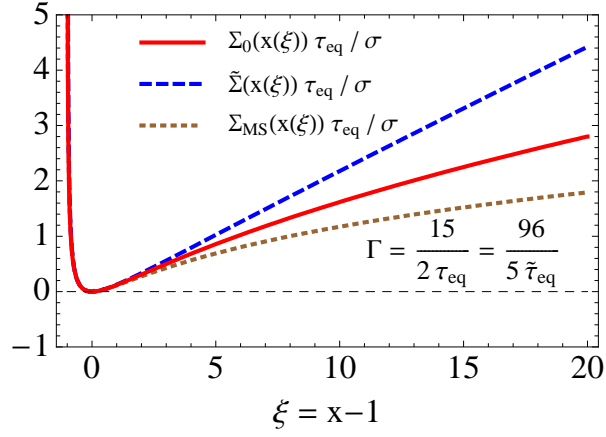}
\end{center}
\caption{Different entropy source terms normalized by $\sigma / \tau_{\rm eq}$. The relaxation times are connected by the formula (\ref{Gammataueq}).}
\label{fig:Sigma}
\end{figure}

\subsection{Martinez and Strickland model}
\label{subsect:MSmodel}

A similar model to ADHYDRO has been introduced recently by Martinez and Strickland ({\rm MS}) \cite{Martinez:2010sc,Martinez:2010sd}. This model is based on the consideration of the two first moment of the Boltzmann kinetic equation with the collision term treated in the relaxation time approximation. The inverse of the relaxation time is denoted by $\Gamma$. 

In this approach the evolution of matter is described by two coupled ordinary differential equations for the momentum anisotropy $\xi$ and the typical hard momentum scale $p_{\rm hard}$. They may be written in the compact form as
\begin{eqnarray}
\frac{d \xi}{d\tau} = \frac{2 (1+ \xi)}{\tau} -\frac{4 \Gamma H_{{\rm MS}}(\xi)}{3},
\label{MSeqn1} \\
\frac{1}{p_{\rm hard}}\frac{d p_{\rm hard}}{d \tau} = \frac{\Gamma}{3} \frac{R^{\prime}_{{\rm MS}} (\xi)}{R_{{\rm MS}} (\xi)} H_{{\rm MS}}(\xi).
\label{MSeqn2}
\end{eqnarray}
The parameter $\xi$ is related to $x$, 
\begin{equation}
\xi = x - 1.
\label{parxi}
\end{equation}
Similarly, the function $R_{{\rm MS}}(\xi)$ is simply connected to $R_0(x)$, 
\begin{eqnarray}
R_{{\rm MS}}(\xi) = \frac{1}{2} \left[ \frac{1}{1+\xi} + \frac{\arctan\sqrt{\xi}}{\sqrt{\xi}}\right]
= \frac{\pi^2}{3 g_0} (1+\xi)^{-2/3} R_0(1+\xi).
\end{eqnarray}
In order to show similarities with our approach, in Eqs. (\ref{MSeqn1}) and (\ref{MSeqn2}) we have introduced the function $H_{{\rm MS}}$ that is an analog of $H_0(x)$,
\begin{eqnarray}
H_{{\rm MS}}(\xi) = \frac{3 (\xi +1) \left(\sqrt{\xi +1} R_{{\rm MS}}^{3/4}(\xi )-1\right) R_{{\rm MS}}(\xi )}{2 R_{{\rm MS}}(\xi )+ 3 (\xi +1) R_{{\rm MS}}'(\xi )}.
\label{Hms}
\end{eqnarray}
Close to equilibrium, the function $H_{{\rm MS}}(\xi(x))$ has the following expansion
\begin{eqnarray}
H_{{\rm MS}}(\xi(x)) \approx \frac{3 }{8}(x-1)+\frac{17}{84} (x-1)^2 + \cdots . \label{expHms} 
\end{eqnarray}
Thus, comparing Eqs. (\ref{xoftau0}) and (\ref{MSeqn1}) we conclude that our standard approach is equivalent to {\rm MS} (near equilibrium) if
\begin{equation}
\Gamma = \frac{15}{2 \tau_{\rm eq} } = \frac{96}{5 {\tilde \tau}_{\rm eq} }.
\label{Gammataueq}
\end{equation}
The plots of the functions $H_0(x(\xi))$, ${\tilde H}(x(\xi))$ and $H_{{\rm MS}}(x(\xi))$, rescaled by the factors defined in (\ref{Gammataueq}), are shown in Fig.~\ref{fig:H}. As it has been already discussed in \cite{Ryblewski:2010bs} the essential difference between these functions is observed only for very large anisotropies.

It is instructive to identify the entropy source in the Martinez and Strickland model. The entropy density in this model has the form \mbox{$\sigma = A p_{\rm hard}^3 x^{-1/2}$}. With the help of Eqs. (\ref{MSeqn1}) and (\ref{MSeqn2}) we reproduce Eq. (\ref{eq3binv}) with
\begin{eqnarray}
\Sigma = \Sigma_{{\rm MS}} &=& \sigma \Gamma 
\left[\frac{R^\prime_{{\rm MS}}(\xi)}{R_{{\rm MS}}(\xi)} +\frac{2}{3(1+\xi)}\right] 
H_{{\rm MS}}(\xi) \nonumber \\ 
&=& \sigma \Gamma \, \frac{R_0'(x)}{R_0(x)} \, H_{{\rm MS}}(\xi(x)) .
\label{en3}
\end{eqnarray}
Thus, comparing Eqs. (\ref{H0}) and (\ref{en3}) we find that {\it the Martinez and Strickland model fits exactly to the ADHYDRO scheme} if the entropy source is defined by (\ref{en3}). 

Moreover, expanding functions $R_{{\rm MS}}(\xi)$ and $H_{{\rm MS}}(\xi)$ for small $\xi$, and using Eq. (\ref{Gammataueq}), we find that $\Sigma_{{\rm MS}}$ agrees with $\Sigma_0$ and ${\tilde \Sigma}$ up to quadratic terms in $\xi$. Different entropy source terms normalized by $\sigma / \tau_{\rm eq}$ are shown in Fig.~\ref{fig:Sigma}. Once again, the relaxation times are identified with the help of Eq. (\ref{Gammataueq}).

\begin{figure}[t]
\begin{center}
\includegraphics[angle=0,width=0.55\textwidth]{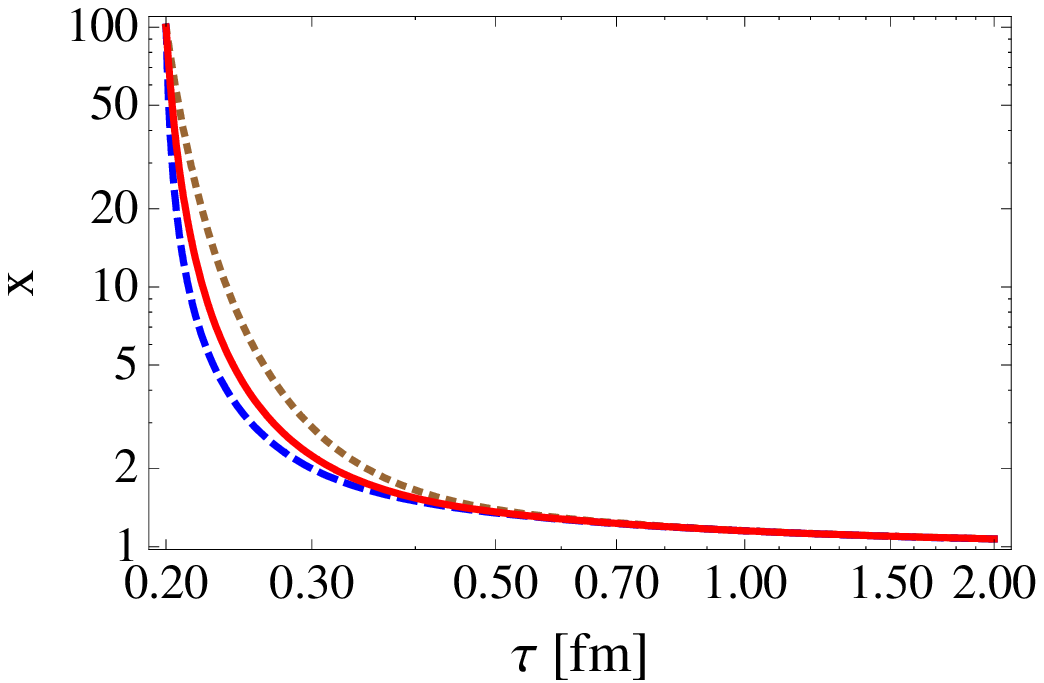} \\
\includegraphics[angle=0,width=0.55\textwidth]{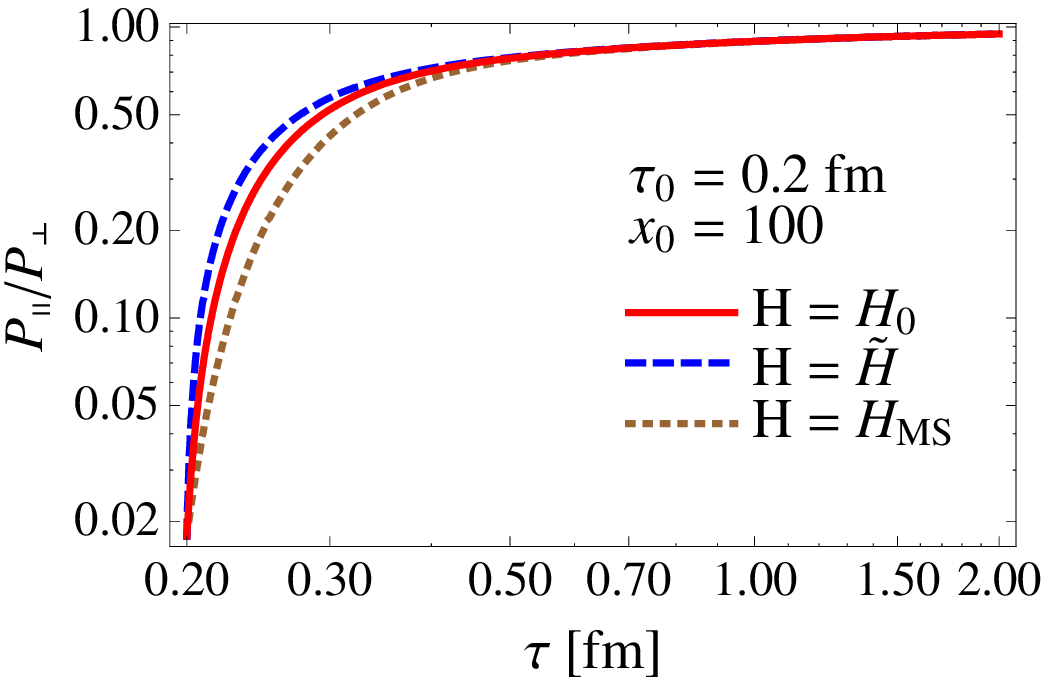} \\
\includegraphics[angle=0,width=0.55\textwidth]{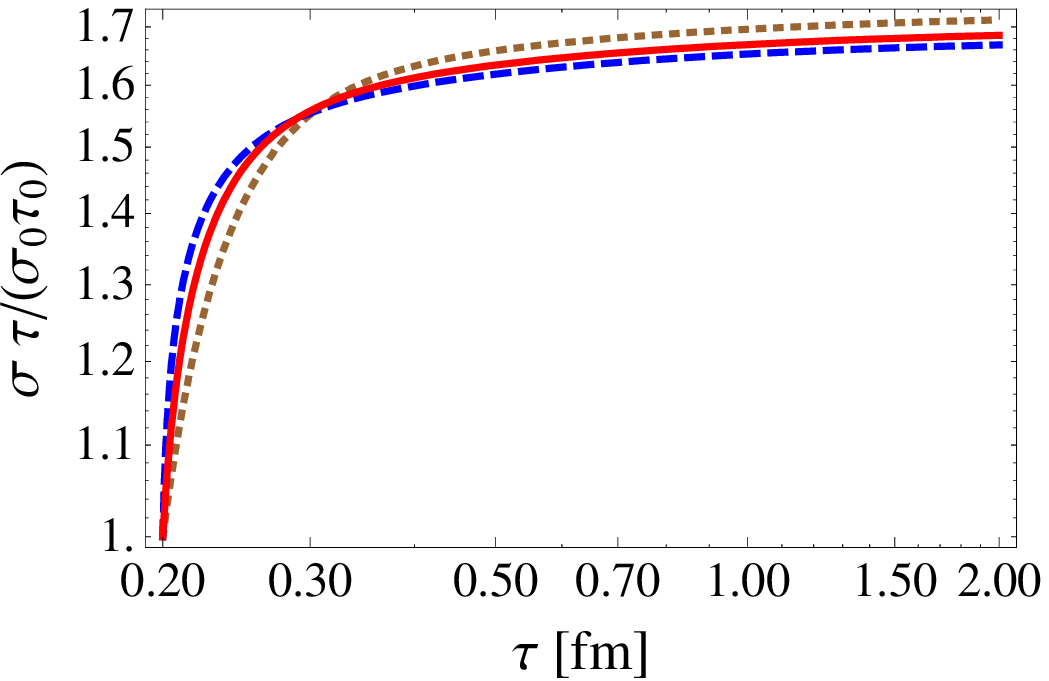} 
\end{center}
\caption{Time dependence of the asymmetry parameter $x$ (upper part), ratio of the longitudinal and transverse pressures (middle part), and entropy density normalized to the values obtained from the Bjorken model, $\sigma_{\rm Bj} = \sigma_0 \tau_0/\tau$  (lower part). The standard ADHYDRO calculations are shown as solid lines, the results of ADHYDRO with the modified entropy source are represented by dashed lines, and the results for the {\rm MS} model are shown as dotted lines.
}
\label{fig:BIresults}
\end{figure}

\section{Results for the boost-invariant case}
\label{subsect:res1}

In this Section we present numerical solutions of Eqs. (\ref{xoftau0}), (\ref{xoftauT}) and (\ref{MSeqn1}). As discussed above, these equations correspond to three different forms of the entropy source defined by Eqs. (\ref{en1}), (\ref{en2}), and (\ref{en3}),  respectively. The parameter $\tau_{\rm eq}$ is set equal to $0.25$ fm. Using this value we find that the system equilibrates at the proper time of about 1 fm. The values of the time-scale parameter ${\tilde \tau_{\rm eq}}$ and the inverse relaxation time $\Gamma$ are connected with $\tau_{\rm eq}$ by Eq. (\ref{Gammataueq}).

In all the cases the initial conditions are specified  at the initial proper time $\tau_0 = 0.2$ fm. We assume that the transverse pressure dominates over the longitudinal pressure during the very early stages of the evolution of matter and take $x(\tau_0)=x_0=100$. In the upper part of Fig. \ref{fig:BIresults} we show the time dependence of $x$. The solid line describes the standard ADHYDRO result, with $\Sigma=\Sigma_0$, the dashed line describes the result obtained with the modified entropy source in ADHYDRO, with $\Sigma={\tilde \Sigma}$, and the dotted line describes the time evolution of $x$ in the {\rm MS} model where $\Sigma=\Sigma_{{\rm MS}}$. 

The middle part of Fig. \ref{fig:BIresults} shows the time evolution of the ratio of the longitudinal and transverse pressure, and the lower part of Fig. \ref{fig:BIresults} shows the time evolution of the entropy density normalized to the initial entropy density. The notation is the same as in the upper part of Fig. \ref{fig:BIresults}.

We observe that the time dependence of $x$ is very much similar in the three cases, suggesting that the particular form of the entropy source has small effect on the time evolution (note the double logarithmic scale). This behavior is caused by the initial large entropy production that leads to a fast decrease of $x$. Later, when $x$ becomes closer to unity, all the schemes are equivalent and imply a very similar time development. 

We find, however, that the equilibration of pressures is a little bit delayed in the {\rm MS} model, as compared to the standard version, and this is reflected in a larger entropy production (relative to the initial entropy density). On the other hand, the equilibration in the modified model where $\Sigma={\tilde \Sigma}$ is a little bit accelerated.

\section{Non-boost-invariant evolution}
\label{sect:nonboostinv}

In the case of purely one-dimensional, non-boost-invariant motion we define
\begin{eqnarray}
U^{\mu} = \left(\cosh\vartheta(\tau,\eta),0,0,\sinh\vartheta(\tau,\eta)\right),  \label{Uv}\\
V^{\mu} = \left(\sinh\vartheta(\tau,\eta),0,0,\cosh\vartheta(\tau,\eta)\right),  \label{Vv}
\end{eqnarray}
where $\vartheta(\tau,\eta)$ is the fluid rapidity which depends on proper time and space-time rapidity. Equations (\ref{Uv}) and (\ref{Vv}) guarantee that Eqs. (\ref{UVnorm}) are automatically satisfied. 

Using Eqs. (\ref{Uv}) and (\ref{Vv}) in Eqs.  (\ref{enmomcon}) and (\ref{engrow}) we derive three partial differential equations for $\sigma(\tau,\eta)$, $\vartheta(\tau,\eta)$, and $x(\tau,\eta)$. The structure of those equations has been discussed recently in Ref. \cite{Ryblewski:2010bs} and in this paper we skip details describing their derivation and form. It is important, however, to present here our initial conditions.

\begin{figure}[t]
\begin{center}
\includegraphics[angle=0,width=1\textwidth]{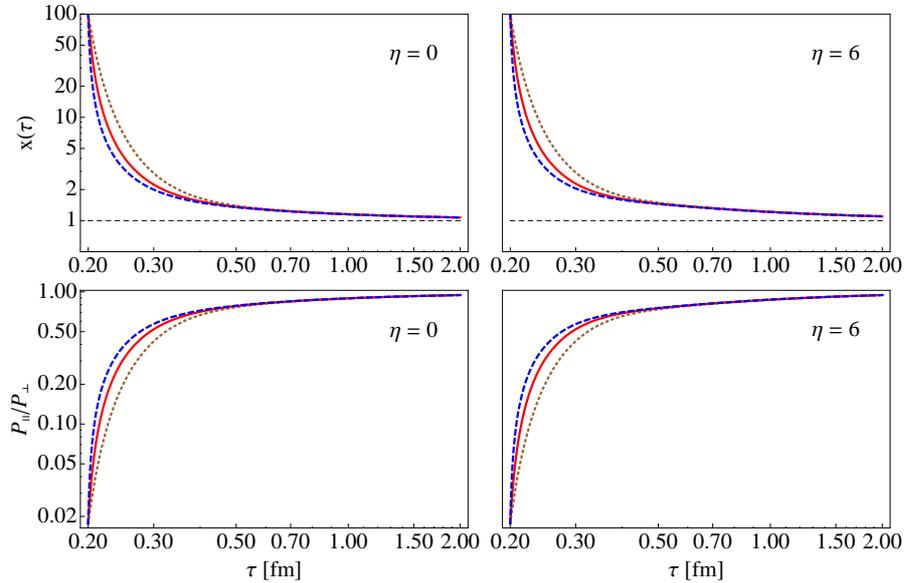}
\end{center}
\caption{Upper parts: time evolution of the anisotropy parameter $x$ for two different values of the space-time rapidity $\eta =0$ (left) and $\eta =6$ (right).  Lower parts: time dependence of the ratio of the longitudinal and transverse pressures, $P_{\parallel}/P_{\perp}$, again for $\eta =0$ (left) and $\eta =6$ (right). The solid lines represent the standard calculation where $\Sigma = \Sigma_0$, the dashed lines show the results with the modified entropy source where $\Sigma = {\tilde \Sigma}$, and the dotted lines show the results for $\Sigma = \Sigma_{{\rm MS}}$.}
\label{fig:NBIresults1}
\end{figure}

\begin{figure}[t]
\begin{center}
\includegraphics[angle=0,width=0.65\textwidth]{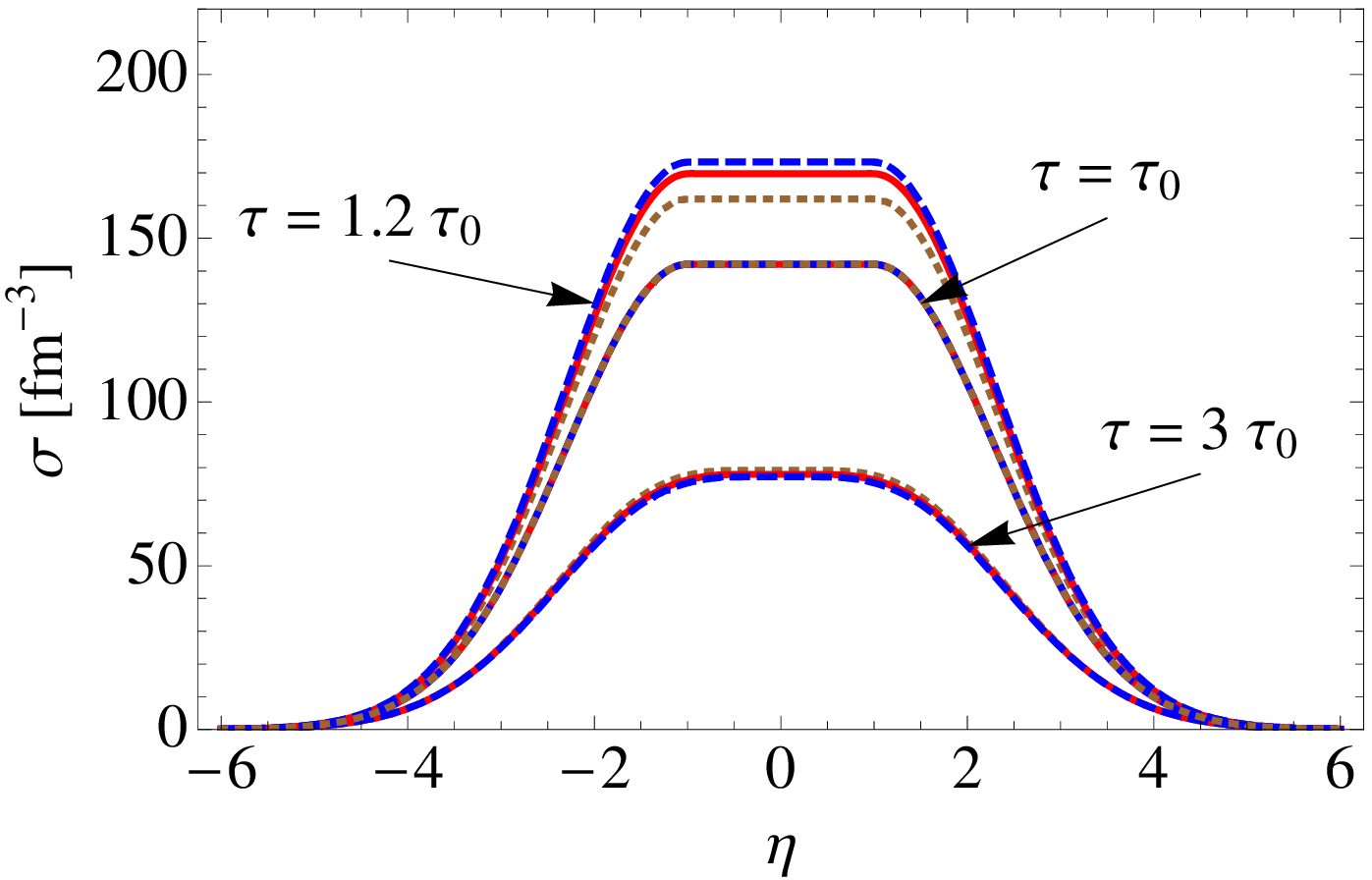}
\includegraphics[angle=0,width=0.65\textwidth]{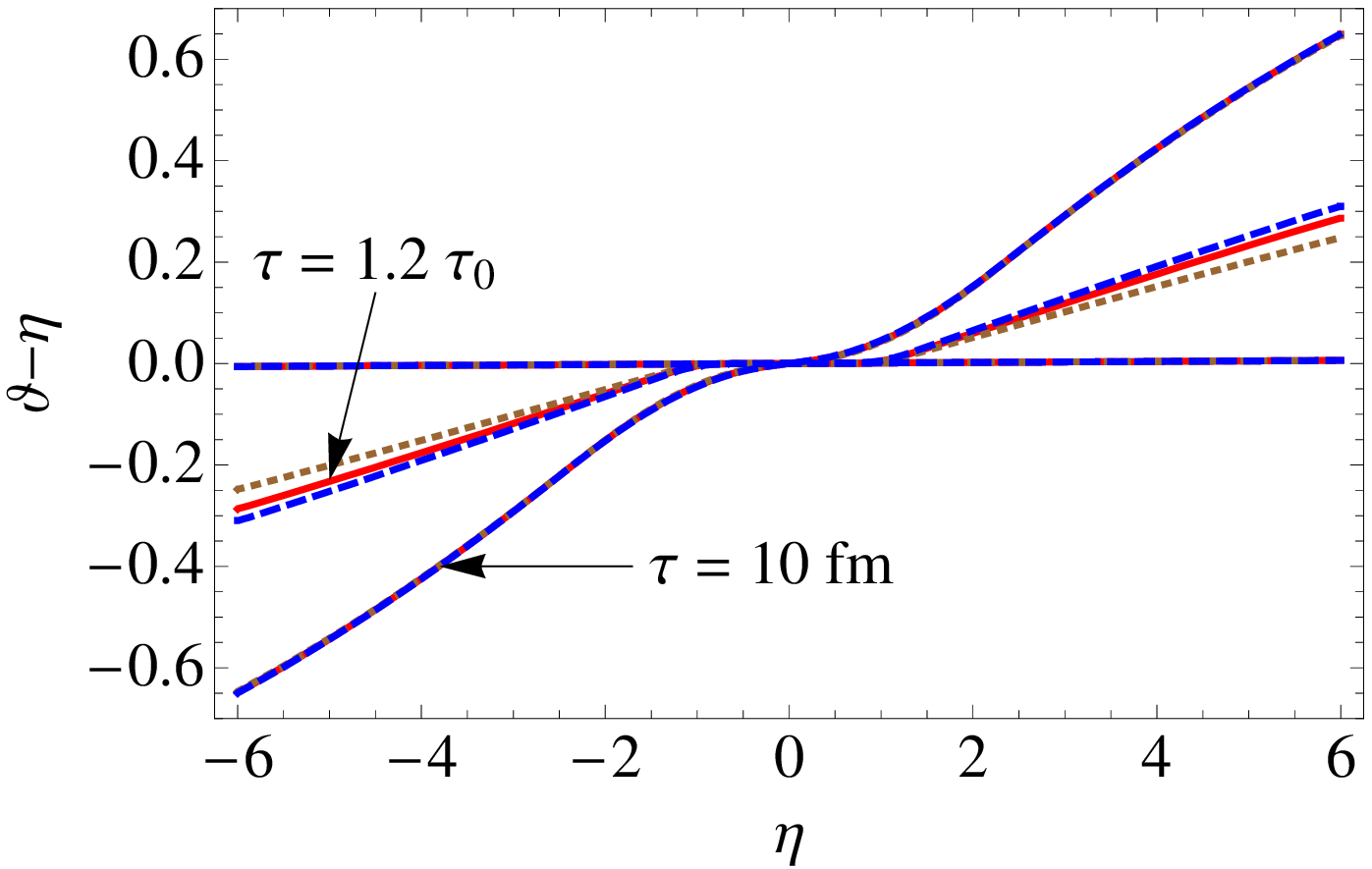}
\end{center}
\caption{Time evolution of the entropy (upper part) and fluid rapidity (lower part) profiles in $\eta$. Similarly as in Fig. \ref{fig:NBIresults1}, the solid, dashed, and dotted lines correspond to three different forms of the entropy source used in the numerical calculations.}
\label{fig:NBIresults2}
\end{figure}

The initial conditions for a non-boost-invariant evolution are defined by the three functions of space-time rapidity: $\sigma(\tau_0,\eta)$, $\vartheta(\tau_0,\eta)$, and $x(\tau_0,\eta)$.  The initial entropy profile is taken in the form \cite{Bozek:2009ty,Hirano:2002ds}
\begin{eqnarray}
\sigma(\tau_0, \eta) = \sigma_0 \exp\left[
-\frac{(|\eta|-\Delta\eta)^2}{2 (\delta\eta)^2} \; \theta(|\eta|-\Delta\eta)
\right],
\label{init_sigma}
\end{eqnarray}
where $\theta$ is the step function, the parameter $\Delta\eta$ defines the half width of the initial  plateau in space-time rapidity, and $\delta\eta$ defines the half width of the Gaussian tails on both sides of the plateau. To match the rapidity distribution measured by BRAHMS \cite{Bearden:2004yx} we use the values: $\Delta\eta=1$ and $\delta\eta=1.3$. The value of the initial central entropy density is obtained from the condition that the initial energy density is 100 GeV/fm$^3$,
\begin{equation}
\varepsilon_0 = 100 \, \hbox{GeV/fm$^3$} = \left(\frac{\pi^2 \sigma_0}{4 g_0} \right)^{4/3} R(x_0).
\label{eps0}
\end{equation}
Here, $x_0$ is the initial value of the anisotropy parameter at $\eta=0$. In this Section we assume that the initial profile of $x$ is constant, 
\begin{equation}
x(\tau_0, \eta) = x_0 = 100.
\label{init_xconst}
\end{equation}
The time-scale parameter $\tau_{\rm eq}$ has the same value as in the previous Section, and the initial fluid rapidity profile is taken in the form
\begin{equation}
\vartheta(\tau_0, \eta) = \eta.
\label{init_vartheta}
\end{equation}

The results of our numerical calculations done with three different forms of the entropy source are shown in Figs. \ref{fig:NBIresults1} and \ref{fig:NBIresults2}. The solid lines represent the standard calculation where $\Sigma = \Sigma_0$, the dashed lines show the results with the modified entropy source where $\Sigma = {\tilde \Sigma}$, and the dotted lines show the results for the case $\Sigma=\Sigma_{{\rm MS}}$. In the two upper parts of Fig. \ref{fig:NBIresults1} we show the time dependence of the anisotropy parameter $x$ for two different values of the space-time rapidity, $\eta =0$ (left) and $\eta =6$ (right). We observe that this dependence is very much similar to the dependence obtained in the boost-invariant calculation, see the upper part of Fig. \ref{fig:BIresults}. Thus the rapidity dependence of the initial entropy profile does not affect the time dependence of $x$ in a noticeable way.  In the two lower parts of Fig. \ref{fig:NBIresults1} we show the time dependence of the ratio of pressures, $P_\parallel/P_\perp$. Again, we observe weak rapidity dependence, compare the middle part of Fig. \ref{fig:BIresults}.

In Fig. \ref{fig:NBIresults2} we show the entropy density $\sigma$ and the fluid rapidity $\vartheta$ as functions of space-time rapidity for the initial proper time $\tau=\tau_0$, a bit later proper time $\tau= 1.2 \,\tau_0$, and $\tau= 3 \, \tau_0$  (for $\sigma$) or $\tau=10$ fm (for $\vartheta-\eta$). At $\tau= 1.2 \,\tau_0$ the entropy profiles are higher than the initial profiles --- at such an early time the entropy is produced in the system. For $\tau= 3 \, \tau_0$ the entropy profiles lie below the initial profiles --- the entropy is not produced in the system anymore, and the entropy density drops due to the longitudinal expansion of the system. We observe that the results obtained with the three forms of $\Sigma$ are very much similar, indicating that the specific form of the entropy form has small effect on the evolution of matter also for the non-boost-invariant systems.

\begin{figure}[t]
\begin{center}
\includegraphics[angle=0,width=0.65\textwidth]{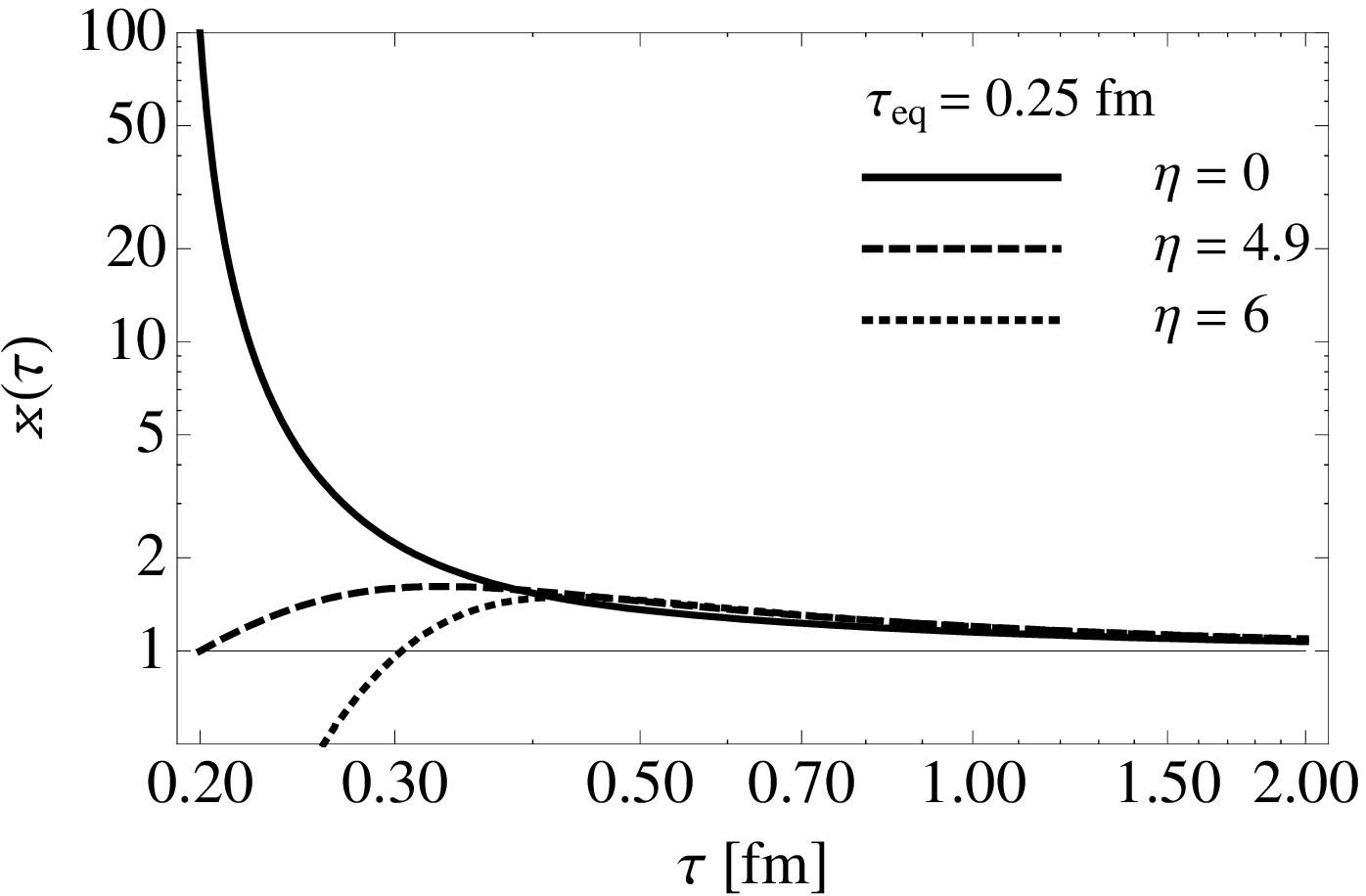}
\includegraphics[angle=0,width=0.65\textwidth]{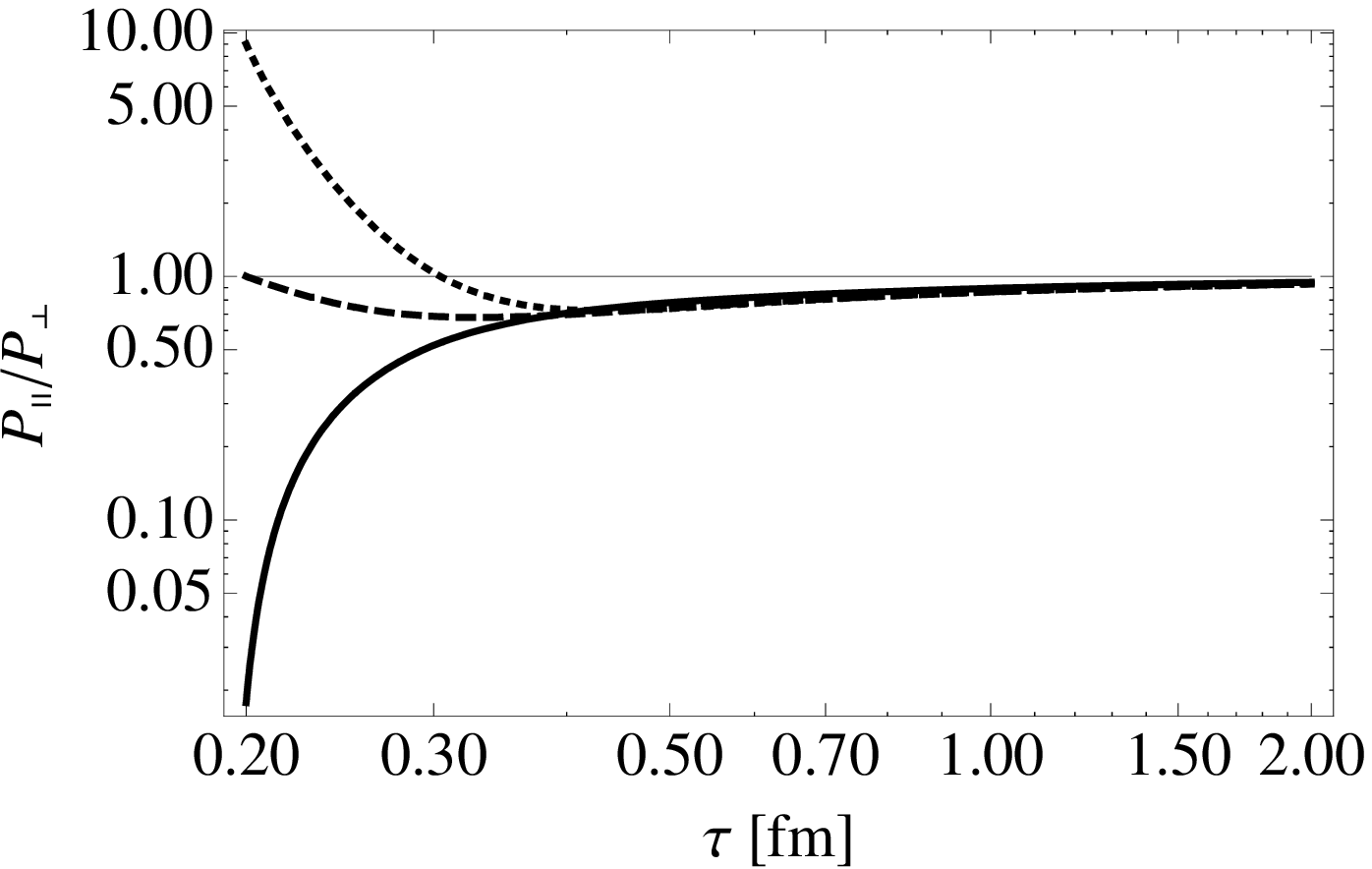}
\end{center}
\caption{Time evolution of the anisotropy parameter $x$ (upper part) and the ratio of pressures (lower part) for three different values of the space-time rapidity: $\eta=0$ (solid lines), $\eta=4.9$ (dashed lines), and $\eta=6$ (dotted lines).
}
\label{fig:NBIAresults1}
\end{figure}

\section{Rapidity dependent anisotropy}
\label{sect:rapidity}
%
In the previous Section we have analyzed the case where the initial anisotropy parameter is independent of space-time rapidity, $x(\tau_0,\eta) = x_0$. In this Section we consider the case where the initial anisotropy profile depends on $\eta$ but other initial conditions are exactly the same. We assume that the initial $\eta$ profile of $x$ is the same as the initial entropy density profile, namely
\begin{eqnarray}
x(\tau_0, \eta) = x_0 \frac{\sigma(\tau_0,\eta)}{\sigma_0},
\label{init_x}
\end{eqnarray}
where $\sigma(\tau_0,\eta)$ is given by Eq. (\ref{init_sigma}). With $x_0=100$ Eq. (\ref{init_x}) implies that the initial anisotropy is large at the center of the system (central region) and small at the edges of the system (fragmentation regions).

In Fig. \ref{fig:NBIAresults1} we show the time evolution of the anisotropy parameter $x$ and the ratio of pressures for three different values of the space-time rapidity: $\eta=0$ (solid lines), $\eta=4.9$ (dashed lines), and $\eta=6$ (dotted lines). The behavior of the system in the central region ($\eta=0$) is very much similar to the behavior of the boost-invariant systems studied before. On the other hand, at large rapidities the time evolution of the physical quantities is quite different, which is caused by a different starting value of $x$. For small initial values of $x$ we observe first the growth of $x$ with time (above unity) and only later a decrease towards unity. Anyway, we observe that at $\tau \approx 0.5$~fm the values of $x$ become practically independent of $\eta$.

\begin{figure}[t]
\begin{center}
\includegraphics[angle=0,width=0.65\textwidth]{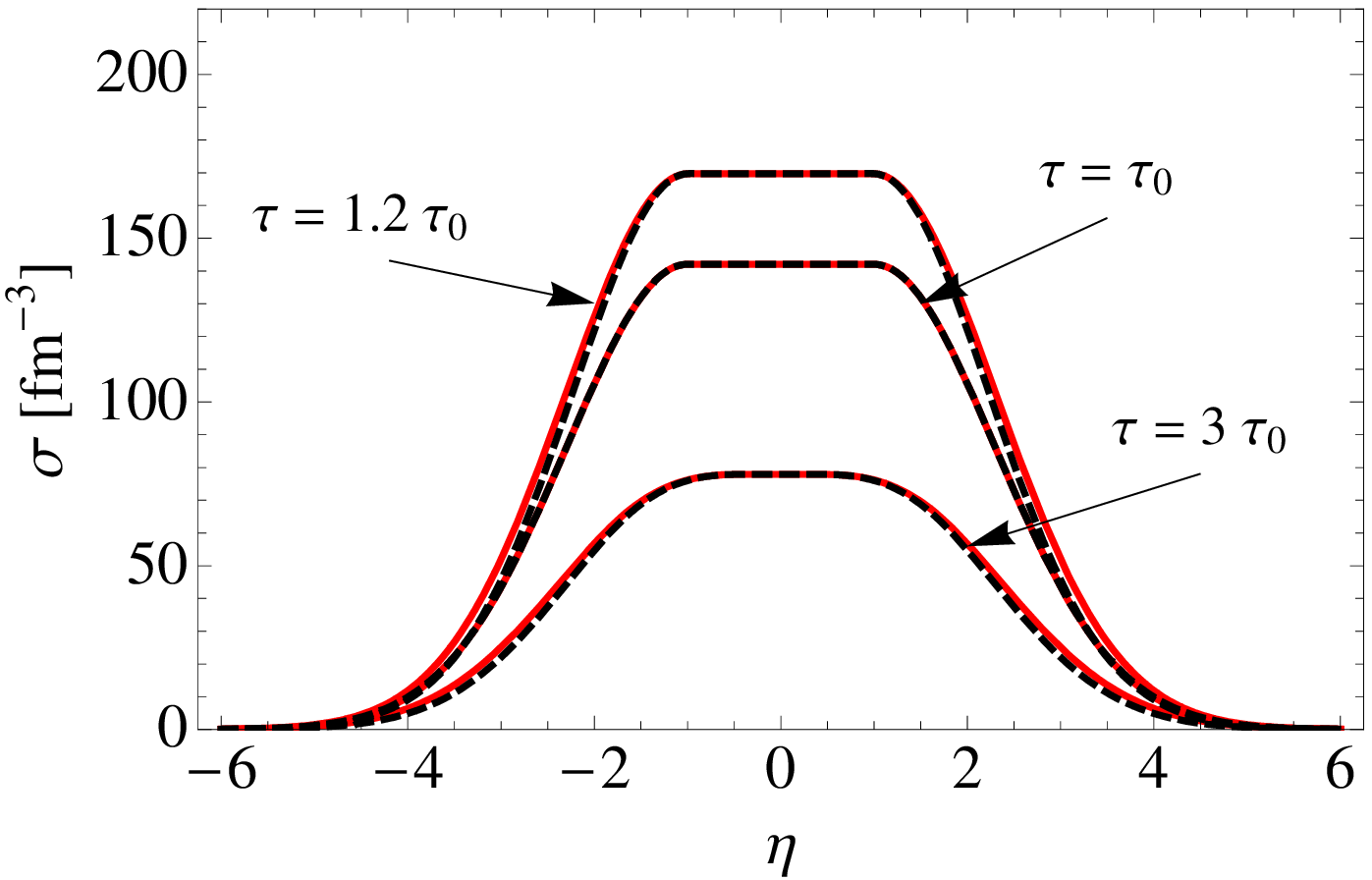}
\includegraphics[angle=0,width=0.65\textwidth]{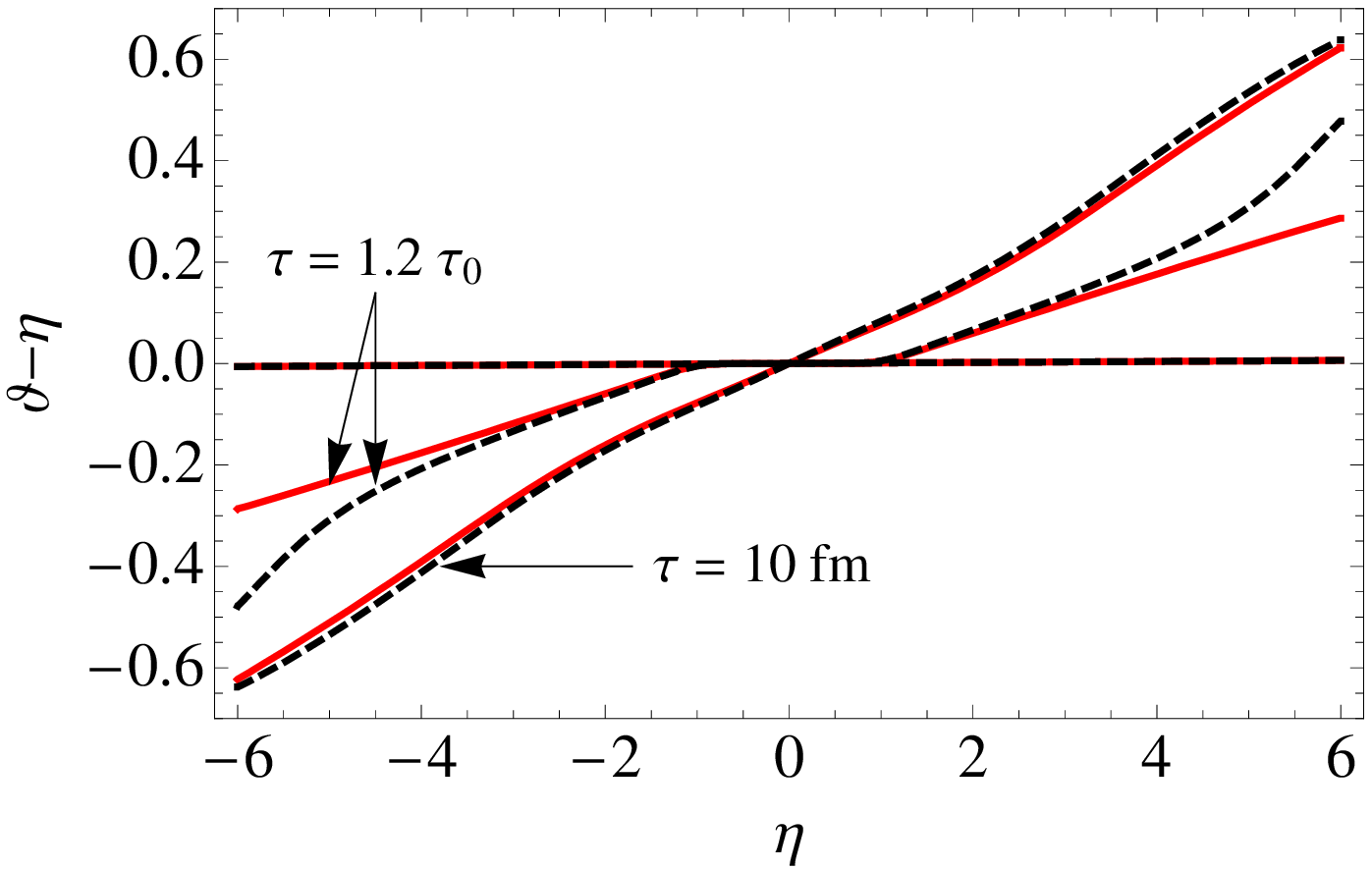}
\end{center}
\caption{Time evolution of the entropy and rapidity profiles for the initial conditions specified by Eqs. (\ref{init_sigma}), (\ref{init_xconst}), and (\ref{init_vartheta}) (solid lines) and by Eqs. (\ref{init_sigma}), (\ref{init_vartheta}) and (\ref{init_x}) (dashed lines).
}
\label{fig:rNBIAesults2}
\end{figure}

In Fig.  \ref{fig:rNBIAesults2} we show the time evolution of the entropy and rapidity profiles for the initial conditions specified by Eqs. (\ref{init_sigma}), (\ref{init_xconst}), and (\ref{init_vartheta}) (solid lines) and by Eqs.  (\ref{init_sigma}), (\ref{init_vartheta}) and (\ref{init_x}) (dashed lines). In the two considered cases the entropy profiles behave in the very similar way. The behavior of the fluid rapidity is different only at the beginning of the evolution for large rapidities. 

\section{Conclusions}
\label{sect:conclusions}

In this paper we have discussed the assumptions of the recently formulated model that may be used to describe the evolution of a highly-anisotropic fluid formed at the very early stages of relativistic heavy-ion collisions. We have considered three forms of the entropy source which defines the dynamics of the system. We have found that the numerical results depend very weakly on the particular form of the entropy source. 

We have also studied microscopic interpretation of the model, discussing different forms of the microscopic phase-space distribution function. We have found that corrections for quantum effects lead to minor quantitative differences as compared to the standard case where the phase space distribution is described by the (stretched or squeezed) Boltzmann distribution. 

We have performed the numerical calculations for the cases where the initial anisotropy depends on space-time rapidity. In such cases the evolution of the system is more complicated but the tendency of the system to equilibrate is not changed.

Altogether, we have found that the results obtained with the ADHYDRO model are stable against various modifications of the model components such as: the entropy source, the form of the microscopic phase-space distribution function, or the form of the initial conditions. We consider such stability as an attractive physical feature of the model proposed in \cite{Florkowski:2010cf}. 

\section{Appendix 1}
\label{sect:app1}

In this Section we give the explicit form of the Lorentz transformation which leads from the center-of-mass frame of the colliding nuclei to the local-rest-frame of the fluid element. This transformation is constructed as a sequence of the longitudinal Lorentz boost, rotation around the $z$-axis, and the transverse Lorentz boost. In this way the complete transformation does not change the longitudinal direction.

The four-vectors $U^\mu$ and $V^\mu$ may be written in the following form
\begin{eqnarray}
U^\mu &=& \cosh\vartheta_\parallel \cosh\vartheta_\perp 
\left( 1, 
\frac{\tanh\vartheta_\perp \cos\phi}{\cosh\vartheta_\parallel}, 
\frac{\tanh\vartheta_\perp \cos\phi}{\cosh\vartheta_\parallel}, 
\tanh\vartheta_\parallel \right), \nonumber \\
V^\mu &=& \cosh\vartheta_\parallel 
\left( \tanh\vartheta_\parallel, 
0, 
0, 
1 \right).
\end{eqnarray}
where $\vartheta_\parallel$ is the longitudinal fluid rapidity and $\vartheta_\perp$ is the transverse fluid rapidity.

The Lorentz transformation leading to the local rest frame consists of the Lorentz boost along the $z$-axis,
\begin{equation}
 L^{\, \mu}_{(z) \, \nu} = \left(
\begin{array}{cccc}
\cosh\vartheta_\parallel & 0 & 0 & -\sinh\vartheta_\parallel \\
0 & 1 & 0 & 0 \\
0 & 0 & 1 & 0 \\
-\sinh\vartheta_\parallel & 0 & 0 & \cosh\vartheta_\parallel
\end{array} \right),
\label{Lz}
\end{equation}
the rotation around the $z$-axis,
\begin{equation}
 R^{\, \mu}_{(xy) \, \nu} = \left(
\begin{array}{cccc}
1 & 0 & 0 & 0 \\
0 & \cos\phi & \sin\phi & 0 \\
0 & -\sin\phi & \cos\phi & 0 \\
0 & 0 & 0 & 1
\end{array} \right) , 
\label{Rxy}
\end{equation}
and the Lorentz boost along the $x$-axis
\begin{equation}
 L^{\, \mu}_{(x) \, \nu} = \left(
\begin{array}{cccc}
\cosh\vartheta_\perp & -\sinh\vartheta_\perp & 0 & 0 \\
-\sinh\vartheta_\perp & \cosh\vartheta_\perp & 0 & 0 \\
0 & 0 & 1 & 0 \\
0 & 0 & 0 & 1
\end{array} \right).
\label{Lx}
\end{equation}
By direct multiplication one may check that
\begin{eqnarray}
 L^{\, \alpha}_{(x) \, \beta} R^{\, \beta}_{(xy) \, \gamma}  L^{\, \gamma}_{(z) \, \mu} U^\mu &=& (1,0,0,0), \nonumber \\
 L^{\, \alpha}_{(x) \, \beta} R^{\, \beta}_{(xy) \, \gamma}  L^{\, \gamma}_{(z) \, \mu} V^\mu &=& (0,0,0,1).
\end{eqnarray}
In order to move from the local rest frame to the center-of-mass frame we perform the inverse transormation.

\section{Appendix 2}
\label{sect:app2}

The concept of ADHYDRO includes, as a special case, the framework of transverse hydrodynamics. In the latter case, we assume that the distribution function (\ref{Fxp2}) has the following factorized form
\begin{eqnarray}
f &=& f_\perp\left( \frac{  \sqrt{(p\cdot U)^2  - (p\cdot V)^2}  }{\lambda_\perp} \right) 
f_\parallel\left( \frac{p\cdot V}{\lambda_\perp} \right) 
= f_\perp\left( \frac{p\cdot U}{\lambda_\perp} \right) 
\frac{n_0}{\tau} \delta\left(p\cdot V \right). \nonumber \\
\label{TH1}
\end{eqnarray}
Here $n_0$ is a normalization constant and $\delta$ is the Dirac delta function. Equation (\ref{TH1}) allows us to make the identification
\begin{equation}
\lambda_\parallel = \frac{1}{\tau}.
\label{TH2}
\end{equation}
Using our general definition of the function $R(x)$, see Eq. (\ref{Rofiks}), we find 
\begin{equation}
R(x) = r x^{1/6}, \quad x R'(x) = \frac{R(x)}{6},
\label{TH3}
\end{equation}
where $r$ is a constant. Equations (\ref{TH3}) used in (\ref{vareR})--(\ref{PLR}) immediately yield
\begin{equation}
P_\perp = \frac{\varepsilon}{2}, \quad P_\parallel = 0.
\label{TH4}
\end{equation}
Hence, the ansatz (\ref{TH1}) corresponds to the situation where the longitudinal pressure vanishes and the system may be treated as a collection of independent clusters which expand transversally. 

The parameter $\lambda_\perp$ describes temperature of two-dimensional transverse clusters. For the purely longitudinal expansion, the parton density drops like $1/\tau$. Since $n \sim \lambda_\perp^2 \lambda_\parallel = \lambda_\perp^2/\tau$, the value of $\lambda_\perp$ remains unchanged. Clearly, only transverse expansion can lower cluster's temperature. Moreover, in such a boost-invariant case $x \sim \tau^2$, hence the ratio of the transverse and longitudinal pressure always grows and the system cannot reach stable isotropic equilibrium. 

\section{Appendix 3}
\label{sect:app3}

In this Appendix we discuss consequences of replacing the classical distribution function by its quantum analogs, the Bose-Einstein and Fermi-Dirac distributions. The integrals over the distribution function (\ref{fBE}) may be done by expanding (\ref{fBE})
in a geometric series,
\begin{equation}
f _{BE} = g_0 \sum_{n=1}^\infty 
\exp\left( -n  \sqrt{p_\perp ^2/\lambda_\perp^2 + 
p_\parallel^2/\lambda_\parallel^2} \, \right), 
\label{fBEexp}
\end{equation}
and by integrating (\ref{fBEexp}) term by term. For example, starting from the general definition of the function $R(x)$, see Eq. (\ref{Rofiks}), we obtain
\begin{eqnarray}
R_{BE}(x) &=& g_0 x^{-1/3} \sum_{n=1}^\infty  \int \frac{d\xi_\perp \xi_\perp d\xi_\parallel}{2\pi^2} 
\sqrt{\xi_\parallel^2 + x \xi_\perp^2} \exp\left( -n  \sqrt{\xi_\perp ^2 + 
\xi_\parallel^2} \, \right). \nonumber \\
\label{FBE1}
\end{eqnarray}
Introducing the new integration variables: $x_\perp = n\, \xi_\perp$ and $x_\parallel = n\, \xi_\parallel$ we find
\begin{eqnarray}
R_{BE}(x) &=& g_0 x^{-1/3} \sum_{n=1}^\infty \frac{1}{n^4} \int \frac{dx_\perp x_\perp dx_\parallel}{2\pi^2} 
\sqrt{x_\parallel^2 + x x_\perp^2}  \exp\left( -  \sqrt{x_\perp ^2 + 
x_\parallel^2} \, \right). \nonumber \\
\label{FBE2}
\end{eqnarray}
The sum over $n$ factorizes and gives the Riemann zeta function $\zeta(4)$. As a consequence, we obtain the simple relation to the ``classical'' result,
\begin{equation}
R_{BE}(x) = \zeta(4) R_0(x) = \frac{\pi^4}{90} R_0(x).
\end{equation}

We note that a similar technique may be used in the calculations involving the Fermi-Dirac distribution. In this case we have
\begin{eqnarray}
f _{FD} &=& \frac{g_0} {\exp 
\left( \sqrt{p_\perp ^2/\lambda_\perp^2 + p_\parallel^2/\lambda_\parallel^2 } \,  \right) 
+ 1} \nonumber \\
&=&  g_0 \sum_{n=1}^\infty (-1)^{n+1}
\exp\left( -n  \sqrt{p_\perp ^2/\lambda_\perp^2 + 
p_\parallel^2/\lambda_\parallel^2} \, \right), 
\label{fFD}
\end{eqnarray}
and Eq. (\ref{Rofiks}) leads us to the expression
\begin{eqnarray}
R_{FD}(x) &=& g_0 x^{-1/3} \sum_{n=1}^\infty  (-1)^{n+1} \int \frac{d\xi_\perp \xi_\perp d\xi_\parallel}{2\pi^2} 
\sqrt{\xi_\parallel^2 + x \xi_\perp^2}  \exp\left( -n  \sqrt{\xi_\perp ^2 + 
\xi_\parallel^2} \, \right). \nonumber \\
\label{FD1}
\end{eqnarray}
As in the standard treatment of the Fermi-Dirac distributions in statistical mechanics, we may change the summation in the following way \cite{LandauSPh}
\begin{eqnarray}
\sum_{n=1,2,3,\ldots} (-1)^{n+1} \cdots &=& \sum_{n=1,3,5,\ldots} \cdots - \sum_{n=2,4,6,\ldots} \cdots
\nonumber \\
&=&  \sum_{n=1,2,3,\ldots} \cdots - \,\, 2 \sum_{n=2,4,6,\ldots} \cdots.
\end{eqnarray}
In the last sum we change $n$ to $2n$ and obtain
\begin{eqnarray}
R_{FD}(x) &=& g_0 x^{-1/3} \sum_{n=1}^\infty  \int \frac{d\xi_\perp \xi_\perp d\xi_\parallel}{2\pi^2} 
\sqrt{\xi_\parallel^2 + x \xi_\perp^2}  \exp\left( -n  \sqrt{\xi_\perp ^2 + 
\xi_\parallel^2} \, \right) \nonumber \\
&& -2 g_0 x^{-1/3} \sum_{n=1}^\infty  \int \frac{d\xi_\perp \xi_\perp d\xi_\parallel}{2\pi^2} 
\sqrt{\xi_\parallel^2 + x \xi_\perp^2}  \exp\left( -2 n  \sqrt{\xi_\perp ^2 + 
\xi_\parallel^2} \, \right). \nonumber \\
\label{FD2}
\end{eqnarray}
Now, in the first integral on the right-hand-side of (\ref{FD2}) we change the integration variables to  $x_\perp = n\, \xi_\perp$ and $x_\parallel = n\, \xi_\parallel$, and in the second integral we change to $x_\perp = 2 n\, \xi_\perp$ and $x_\parallel = 2 n\, \xi_\parallel$. In this way we find
\begin{eqnarray}
R_{FD}(x) &=& \frac{7 g_0}{8} x^{-1/3}   \sum_{n=1}^\infty  \frac{1}{n^4} 
\int \frac{dx_\perp x_\perp dx_\parallel}{2\pi^2} 
\sqrt{x_\parallel^2 + x x_\perp^2}  \exp\left( -  \sqrt{x_\perp ^2 + 
x_\parallel^2} \, \right), \nonumber \\
\label{FD3}
\end{eqnarray}
where the factor of $7/8$ follows from the term $1 - 2^{1-p}$, where $p$ is the total power of $\xi_\perp^{\,\,'}s$ and $\xi_\parallel^{\,\,'}s$ appearing in the second integral in (\ref{FD2}). Thus, the final result is
\begin{equation}
R_{FD}(x) = \frac{7 \zeta(4)}{8} R_0(x) = \frac{7 \pi^4}{720} R_0(x).
\label{FD4}
\end{equation}
We stress that the definition of $R_0(x)$ involves the number of the internal degrees of freedom $g_0$, hence when dealing with both bosons and fermions at the same time, one should introduce $R_0(x)$ separately for bosons and fermions and add them together, 
\begin{equation}
R_{BE+FD}(x) = \frac{\pi^2}{60} \left(g_0^{BE} + \frac{7g_0^{FD}}{8}  \right)
x^{-\frac{1}{3}} \left[ 1 + \frac{x \arctan\sqrt{x-1}}{\sqrt{x-1}}\right].
\label{RBEFD}
\end{equation}
Here $g_0^{BE}$ and $g_0^{FD}$ are the numbers of internal degrees of freedom of bosons and fermions, respectively.

The parameter $g$ defined by Eq. (\ref{gconst}) may be also calculated for the case of massless bosons and fermions. Following the same steps as in the calculation of the functions $R_{BE}(x)$ and $R_{FD}(x)$ we obtain
\begin{equation}
g_{BE} = \zeta(3) \frac{g^{BE}_0}{\pi^2},
\label{gBE}
\end{equation}
\begin{equation}
g_{FD} = \zeta(3) \left(1-2^{-2} \right) \frac{g^{FD}_0}{\pi^2},
\label{gBE}
\end{equation}
and for the sum
\begin{equation}
g_{BE+FD} = \frac{\zeta(3)}{\pi^2} \left( g^{BE}_0 + \frac{3 g^{FD}_0}{4} \right).
\label{gBEFD}
\end{equation}
Equations (\ref{RBEFD}) and (\ref{gBEFD}) may be used for $R(x)$ and $g$ appearing in Eqs. (\ref{vareR})--(\ref{PLR}) if the system of quarks and gluons with the same anisotropy parameter $x$ and the total density $n$ is described.


\end{document}